\documentclass[twocolumn,prb,showpacs,superscriptaddress]{revtex4}
\usepackage{graphicx}
\usepackage{amsmath}

\begin{document}

\title{Direct observation of electron doping in
La$_{0.7}$Ce$_{0.3}$MnO$_3$ using x-ray absorption spectroscopy}

\author{C. Mitra}
\altaffiliation{present address: Dept. Materials Science,
University of Cambridge, Pembroke St, Cambridge CB2 3QZ, UK}
\affiliation{Max Planck Institute for Chemical Physics of Solids,
N\"othnitzer Str. 40, 01187 Dresden, Germany}
\author{Z. Hu}
\email{zhiwei@ph2.uni-koeln.de} \affiliation{II. Physikalisches
Institut, Universit\"at zu K\"oln, Z\"ulpicher Str. 77, 50937
K\"oln, Germany}
\author{P. Raychaudhuri}
\affiliation{School of Physics and Astronomy, University of
Birmingham, Edgbaston, Birmingham, B15 2TT, UK}
\author{S. Wirth}
\affiliation{Max Planck Institute for Chemical Physics of Solids,
N\"othnitzer Str. 40, 01187 Dresden, Germany}
\author{S.~I. Csiszar}
\affiliation{Materials Science Centre, University of Groningen,
Nijenborgh 4, 9747 AG Groningen, The Netherlands}
\author{H.~H. Hsieh}
\affiliation{Synchrotron Radiation Research Center, Hsinchu 30077,
Taiwan}
\author{H.-J. Lin}
\affiliation{Synchrotron Radiation Research Center, Hsinchu 30077,
Taiwan}
\author{C.~T. Chen}
\affiliation{Synchrotron Radiation Research Center, Hsinchu 30077,
Taiwan}
\author{L.~H. Tjeng}
\affiliation{II. Physikalisches Institut, Universit\"at zu K\"oln,
Z\"ulpicher Str. 77, 50937 K\"oln, Germany}

\date{\today}

\begin{abstract}
We report on a X-ray absorption spectroscopic (XAS) study on a
thin film of La$_{0.7}$Ce$_{0.3}$MnO$_3$, a manganite which was
previously only speculated to be an electron doped system. The
measurements clearly show that the cerium is in the Ce(IV) valence
state and that the manganese is present in a mixture of Mn$^{2+}$
and Mn$^{3+}$ valence states. These data unambiguously demonstrate
that La$_{0.7}$Ce$_{0.3}$MnO$_3$ is an electron doped colossal
magnetoresistive manganite, a finding that may open up new
opportunities both for device applications as well as for further
basic research towards a better modelling of the colossal
magnetoresistance phenomenon in these materials.
\end{abstract}
\maketitle

Hole doped rare-earth manganites of the form R$_{1-x}$A$_x$MnO$_3$
(R = rare-earth, A = divalent cation) have been at the focus of
attention in recent times in the field of ferromagnetic
oxides.\cite{jin,coey} The interest in these compounds stems from
a variety of reasons. These compounds exihibit a large
magnetoresistance, coined as `colossal magnetoresistance' (CMR),
close to their ferromagnetic transition temperature ($T_c$) which
makes them potential candidates for device applications. These
materials exhibit a strong interplay between spin, charge and
orbital degrees of freedom, due to the competition of the various
relevant energy scales that are of comparable magnitude. All these
give rise to a wide variety of phenomena such as electronic phase
separation, charge ordering, spin glass order and half
metallicity.\cite{coey} It is therefore of prime fundamental
interest to study the rich phase diagram of these compounds as a
function of doping $x$ and size of the R/A cation.\cite{schi}

The basic physics of the hole doped rare-earth manganites can be
understood from an interplay of a strong Hund's rule coupling in
the manganese and the Jahn-Teller distortion.\cite{mil} The
divalent cation in these compounds brings the manganese from a
Mn$^{3+}$ valence state in the parent compound into a mixture of
Mn$^{3+}$ and Mn$^{4+}$. Double exchange between the Mn$^{3+}$ and
the Mn$^{4+}$ cations drives the insulating antiferromagnetic
ground state present in the parent compound (LaMnO$_3$) into a
ferromagnetic metallic ground state for $x \gtrsim 0.2$ in the
doped compound. Above $T_c$, for compounds such as
La$_{0.7}$Ca$_{0.3}$MnO$_3$, the decrease in mobility of the
electrons due to spin disorder localizes the carriers via the
formation of Jahn-Teller polarons. This local deformation around
the Jahn-Teller cation Mn$^{3+}$ gives rise to a polaronic
insulating state.\cite{jak} At temperatures close to $T_c$, i.e.
at the transition between the two phases, an external magnetic
field can give rise to CMR. A salient question to ask in this
context is: {\em Can one induce a ferromagnetic metallic ground
state by doping electrons instead of holes in the parent compound
LaMnO$_3$?} Electron doping is expected to drive the manganese
into a mixture of Mn$^{2+}$ and Mn$^{3+}$. This question is
particularly relevant in manganites since manganese can exist in
many valence states. In addition, the system has an intrinsic
symmetry since Mn$^{2+}$ and Mn$^{4+}$ are both non Jahn-Teller
ions whereas Mn$^{3+}$ is a Jahn-Teller ion. Thus, the basic
physics in terms of Hund's rule coupling and Jahn-Teller effect
could operate in the electron doped phase as well. The existence
of an electron doped manganite is not merely of academic interest
since it also opens up possibilities for fabricating novel bipolar
devices using the electron and hole doped manganites where both
spin and charge are utilized.

The crucial issue that we have to address now is {\em whether it
is really possible to electron dope LaMnO$_3$.} One compound,
which displays properties remarkably similar to
La$_{0.7}$Ca$_{0.3}$MnO$_3$, is the cerium doped manganite
La$_{0.7}$Ce$_{0.3}$MnO$_3$. La$_{0.7}$Ce$_{0.3}$MnO$_3$ has a
ferromagnetic metallic ground state with $T_c \sim$ 250
K.\cite{man,ray,mit1} The ferromagnetic transition is accompanied
by a metal-insulator transition and the system has a
magnetoresistance ($\rho (0) - \rho (H) / \rho (0)$) in excess of
70\% at a field of 1.5 T.\cite{ray} It has been shown very
recently that a tunnel junction made of the hole doped
La$_{0.7}$Ca$_{0.3}$MnO$_3$ and La$_{0.7}$Ce$_{0.3}$MnO$_3$
exhibits rectifying characteristics \cite{mit2} in the polaronic
insulating state at temperatures $T > T_c$. While this result may
suggest that cerium doping drives the manganese in a mixture of
Mn$^{2+}$ and Mn$^{3+}$ valencies, there has been so far no direct
evidence for electron doping in this compound. One of the main
difficulty in this compound is that the system forms in {\em
single phase} only in the epitaxial thin film form,\cite{mit1}
deposited through the energetic pulsed laser ablation process
(other methods failed in this respect\cite{joy}). This precludes
the possibility of using conventional techniques such as chemical
analysis. In addition, it is {\it a priori} not obvious that
cerium has enough reducing power to drive Mn$^{3+}$ towards the
electro positive Mn$^{2+}$, in view of the observation that
formally tetravalent CeO$_2$ has an extremely large amount of
oxygen holes.\cite{Hu99}

In this paper we provide a direct evidence of the Ce and Mn
valence states in La$_{0.7}$Ce$_{0.3}$MnO$_3$ through X-ray
absorption measurements on epitaxial thin films of these
compounds.

Bulk polycrystalline targets of La$_{0.7}$Ce$_{0.3}$MnO$_3$ and
LaMnO$_3$ were prepared by a solid state reaction route as
reported earlier.\cite{mit1} Using these targets, epitaxial films
were deposited by pulsed laser deposition (PLD) on SrTiO$_3$ (STO)
substrate, using a KrF excimer laser. The substrate temperature
was kept between 790$^o$C and 800$^o$C at all times. The laser
energy density was approximately 3 J/cm$^2$ with a repetition rate
of 9 Hz and the laser wavelength was 248 nm. The
La$_{0.7}$Ce$_{0.3}$MnO$_3$ film was grown at an oxygen pressure
of 400 mTorr whereas the LaMnO$_3$ film was deposited at 100
mTorr. The lower oxygen pressure during growth of the LaMnO$_3$
film was necessary to avoid over-oxygenation which can induce
considerable hole doping in this compound. After deposition, the
laser ablation chamber was vented with high purity oxygen and the
substrate cooled down to room temperature. Details of the x-ray
characterization can be found in Ref.~\onlinecite{mit1}. Several
reflections were taken in the x-ray analysis to ascertain the
single phase nature of the films.

Magnetization was measured using a Quantum Design superconducting
quantum interference device (SQUID). The LaMnO$_3$ had a weak
\begin{figure}[tb]
\centering \includegraphics[width=8.4cm]{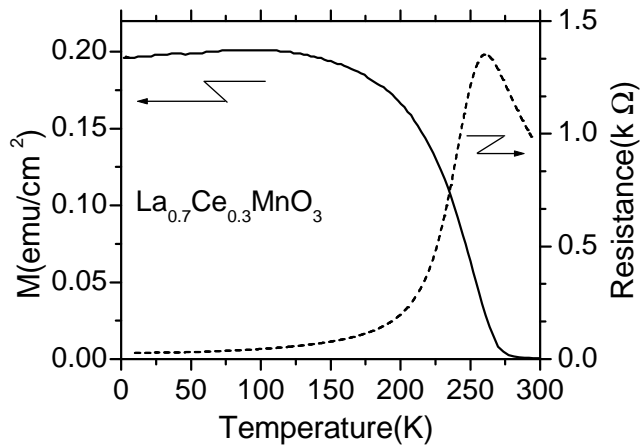}
\caption{Temperature dependence of the magnetization (full line,
left scale) and resistance (dashed line, right scale) of the
investigated La$_{0.7}$Ce$_{0.3}$MnO$_3$ thin film sample. The
magnetization measurement was conducted in an applied field of 1
kOe.} \label{MRT}
\end{figure}
ferromagnetic transition around 115 K. This weak ferromagnetism
could be due to a slight oxygen non-stoichiometry in the film as
has been observed before in bulk samples.\cite{toe} Also, a weak
ferromagnetic moment in the A-type antiferromagnet LaMnO$_3$ is
believed to arise from the Dzyaloshinsky-Moryia
interaction.\cite{coey} The temperature dependence of the
magnetization for the film of interest,
La$_{0.7}$Ce$_{0.3}$MnO$_3$, is shown in Fig.~\ref{MRT}. A $T_c
\approx$ 260 K can be inferred. From the magnetization data there
is no sign of any magnetic impurity phase in our sample. In
addition, the hallmark of the CMR material is depicted in
Fig.~\ref{MRT}: a maximum of the resistance in dependence on
temperature near $T_C$. There is a similarity in the magnetic
properties of La$_{0.7}$Ce$_{0.3}$MnO$_3$ and its hole doped
analogue La$_{0.7}$Ca$_{0.3}$MnO$_3$ in the sense that both have
similar Curie temperatures. It is known that perovskite rare-earth
manganites such as LaMnO$_3$ can accept a large excess of oxygen
via the formation of cation vacancies inducing hole doping in the
parent compound.\cite{toe} Therefore, to confirm that the system
La$_{0.7}$Ce$_{0.3}$MnO$_3$ is indeed electron doped and the
magnetic ordering does not occur due to the existence of Mn$^{4+}$
and Mn$^{3+}$ valence states resulting from over-oxygenation of
the film one must study both the Ce as well as the Mn valence
states.

The Ce-$M_{4,5}$ and Mn-$L_{2,3}$ XAS measurements were performed
using the Dragon beamline at the Synchrotron Radiation Research
Center (SRRC) in Taiwan. The experimental photon energy resolution
was approx. 0.2 and 0.3 eV at the Mn-$L_{2,3}$ and Ce-$M_{4,5}$
thresholds, respectively. The spectra were recorded at room
temperature, employing the total electron-yield method by
measuring the sample drain current to ground. In addition, the
incident beam flux was monitored simultaneously with a gold mesh
in the beam line which allowed the obtained spectra to be
normalized very accurately.

\begin{figure}[tb]
\centering \includegraphics[width=7.8cm]{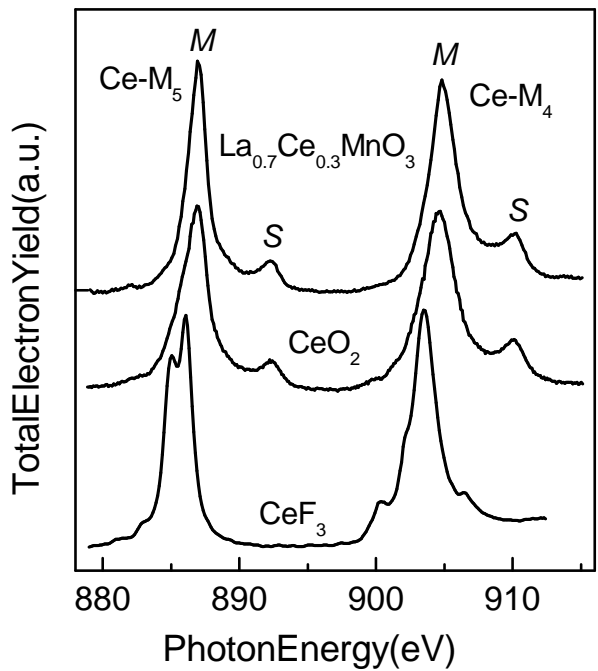}
\caption{Ce-$M_{4,5}$ XAS spectra of La$_{0.7}$Ce$_{0.3}$MnO$_3$
and of CeO$_2$ for a tetravalent and CeF$_3$ for a trivalent
reference.} \label{fig2}
\end{figure}

It is well known that the x-ray absorption spectra at the
rare-earth $M_{4,5}$ and the 3d transition metal $L_{2,3}$
thresholds are highly sensitive to the valence state and the
distribution of valence electrons between the metal ion and the
ligand orbitals since the experimental spectral structures can be
well reproduced by atomic multiplet
calculations.\cite{tho,kot,gro,gro2,memorial} Fig.~\ref{fig2}
shows the Ce-$M_{4,5}$ XAS spectrum of
La$_{0.7}$Ce$_{0.3}$MnO$_3$. For comparison, the spectra of the
reference materials CeO$_2$ and CeF$_3$ are also included. The
Ce-$M_{4,5}$ XAS spectrum of the formally tetravalent Ce-compound
CeO$_2$ consists of a single main structure ($M$) at 887 eV and a
satellite ($S$) at $\cong$ 6 eV above the main peak. The
tetravalence of the Ce in La$_{0.7}$Ce$_{0.3}$MnO$_3$ can be
easily recognized since the spectral features are very similar to
those of CeO$_2$. In tetravalent compounds like CeO$_2$, the
Ce--4$f$ and O--2$p$ covalence can be described\cite{kot} in a
configuration interaction scheme, in which the ground state is
given by $| g\rangle = \alpha \, |\, 4f^0\rangle + \beta \, |\,
4f^1 \underline{L} \rangle$ ($\underline{L}$ refers to the ligand
hole) while the final states can be labelled as the bonding and
anti-bonding states of $3d^9 4f^1$ and $3d^9 4f^2 \underline{L}$.
These two final states give rise to the main peak and the higher
energy satellite. By an interference effect\cite{kot} the
intensity of the anti-bonding state is strongly reduced,
especially if the charge transfer energy between the two
configurations in the ground state and the final state is nearly
the same. Therefore, the higher energy anti-bonding state appears
usually as a weak satellite. Nevertheless, the presence of such a
weak satellite indicates that the $4f$ occupancy is about 0.5, as
a careful theoretical analysis has shown.\cite{kot}

In contrast to the single-peaked structure in Ce(IV)-compounds,
the main spectral features of the trivalent Ce-compound CeF$_3$
exhibits a more complicated multiplet structure which could be
very well reproduced by atomic multiplet calculations.\cite{tho}
It is well known \cite{hu3} that the $M_{4,5}$ spectrum of
trivalent rare-earth elements is shifted to lower energy by more
than 1 eV with a decrease in their valence. Here, we observe a
shift of $\cong 1.5$ eV towards lower energy in the CeF$_3$
spectrum with respect to the spectra of CeO$_2$ and
La$_{0.7}$Ce$_{0.3}$MnO$_3$. Therefore, the Ce-$M_{4,5}$ spectrum
of Ce(IV)-compounds is very sensitive to trivalent impurities. One
can see that even CeO$_2$ has some very small Ce$^{3+}$ content as
suggested by the small asymmetry in the two peaks near the low
energy side. The main peaks of the Ce $M_{4,5}$-edge in
La$_{0.7}$Ce$_{0.3}$MnO$_3$, however, are perfectly symmetric,
indicating a pure Ce(IV) valence state of Ce in this compound.

\begin{figure}[tb!]
\centering \includegraphics[width=7.8cm]{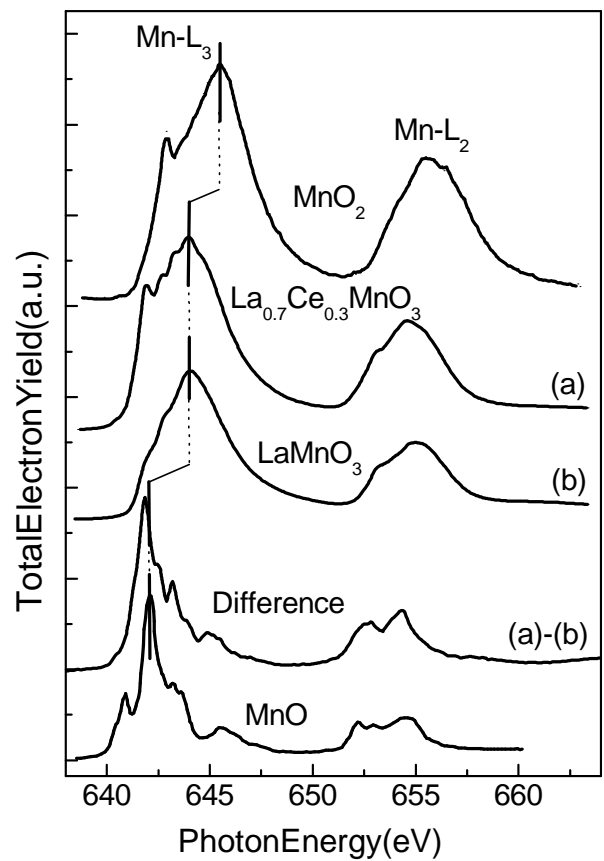}
\caption{Mn-$L_{2,3}$ XAS spectra of La$_{0.7}$Ce$_{0.3}$MnO$_3$
(a) and for comparison of MnO$_2$, LaMnO$_3$ (b) and MnO for
Mn$^{4+}$, Mn$^{3+}$ and Mn$^{2+}$ references, respectively. The
curve labeled ``difference" ((a)-(b)) is the difference between
the La$_{0.7}$Ce$_{0.3}$MnO$_3$ and the LaMnO$_3$ spectra. In
comparison to the MnO spectrum, this difference curve clearly
indicates the existence of Mn$^{2+}$ in
La$_{0.7}$Ce$_{0.3}$MnO$_3$.} \label{fig3}
\end{figure}

The existence of Ce(IV), however, still does not confirm that the
sample is electron doped. In the past it has been shown that in a
bulk polycrystalline sample some amount of CeO$_2$ remains
unreacted though the epitaxial thin film of
La$_{0.7}$Ce$_{0.3}$MnO$_3$ forms in single phase.\cite{mit1}
Thus, to conclusively establish that it is indeed an electron
doped system one has to search for a corresponding replacement of
Mn$^{3+}$ by Mn$^{2+}$ as well. We have measured the Mn-$L_{2,3}$
XAS spectra to investigate the valence state of Mn in the ground
state. Fig.~\ref{fig3} shows the Mn-$L_{2,3}$ spectra of
La$_{0.7}$Ce$_{0.3}$MnO$_3$ and---for comparison---of MnO$_2$,
LaMnO$_3$ and MnO for Mn$^{4+}$, Mn$^{3+}$ and Mn$^{2+}$
references, respectively. The LaMnO$_3$ spectrum is similar to
that obtained previously.\cite{abb,cas} It is a well known
fact\cite{ctchen,hu1,hu2} that an increase of the metal ion
valence by one results in shift of the $L_{2,3}$ XAS spectra to
higher energy by about 1 eV or more. In Fig.~\ref{fig3} we can see
a shift towards higher energy from bottom to top in a sequence of
increasing Mn valence from Mn$^{2+}$ (MnO) to Mn$^{3+}$
(LaMnO$_3$) and further to Mn$^{4+}$ (MnO$_2$). In comparison to
undoped LaMnO$_3$, in La$_{0.7}$Ce$_{0.3}$MnO$_3$ we can see new
and sharp low energy structures at nearly the same energy position
as in the MnO spectrum. These sharp structures at around 642 eV
are a reliable benchmark of the appearance of a divalent Mn state
since they are hardly smeared by background or other structures.
The observed spectral features indicate the existence of a
Mn$^{2+}$ component in addition to Mn$^{3+}$ in the single phase
La$_{0.7}$Ce$_{0.3}$MnO$_3$ compound. In order to estimate the
Mn$^{2+}$ content, the {\em normalized} spectrum of LaMnO$_3$ has
been substracted from that of La$_{0.7}$Ce$_{0.3}$MnO$_3$ (the
resultant difference spectrum is labeled ``difference" in
Fig.~\ref{fig3}). The most reasonable difference spectrum was
obtained if the spectral weight of the substracted LaMnO$_3$
spectrum was taken as 82\%. The main structures of the difference
spectrum are found at the same energy position as the prominent
features of the MnO spectrum (as indicated by the vertical line in
Fig.~\ref{fig3}) and the overall appearance of the difference and
the MnO spectra is very similar. However, the difference spectrum
in its details is not exactly the same as that of MnO. As an
example, there is a sharp shoulder (at around 641 eV) below the
main peak found in MnO which is nearly absent in the difference
spectrum. Such subtle distinctions are to be expected because of
the difference in the local symmetry of Mn in these two compounds.
In fact, there is an additional feature just above the main peak
in the $L_3$-edge of the difference spectrum and the main peak is
further shifted to lower energy by 0.2 eV. Also, a change at the
Mn $L_2$-edge is clearly seen.

In order to estimate the contents of Mn$^{2+}$ and Mn$^{3+}$ from
the corresponding spectral weights (18\% and 82\%, respectively)
one needs to bear in mind that, formally, Mn$^{2+}$ has 5 holes in
the 3d-orbital whereas Mn$^{3+}$ has 6 holes. Assuming completely
ionic Mn$^{2+}$ and Mn$^{3+}$, the content of Mn$^{2+}$ is about
20\%. However, the strong covalence of Mn$^{3+}$ may reduce this
number slightly. Even though the latter effect is difficult to
account for the main error originates from the uncertainty in
determining the spectral weight of Mn$^{2+}$ and hence, we
estimate the Mn$^{2+}$-content in La$_{0.7}$Ce$_{0.3}$MnO$_3$ to
20($\pm 4$)\%. The deviation from the nominal 30\% Ce doping may
be due to the fact that the system might have been over-oxygenated
which gives rise to an excess of Mn$^{3+}$-content.

In summary, we have shown from XAS measurements that
La$_{0.7}$Ce$_{0.3}$MnO$_3$ is an electron doped manganite where
the formally tetravalent cerium drives the manganese into a
mixture of Mn$^{2+}$ and Mn$^{3+}$ valence states. Thus, it is
tempting to think that the ferromagnetism in this compound is
brought about by the double exchange between Mn$^{2+}$/Mn$^{3+}$
ions. The remarkable similarity between this compound and its hole
doped counterpart, La$_{0.7}$Ca$_{0.3}$MnO$_3$, provides further
motivation to explore the electron doped phase in detail. One
issue of particular interest is the spin state of manganese in
La$_{0.7}$Ce$_{0.3}$MnO$_3$. Recent observation of a large
positive magnetoresistance in ferromagnetic tunnel junctions
involving La$_{0.7}$Ce$_{0.3}$MnO$_3$ and a hole doped manganite
as the ferromagnetic electrodes suggests that the manganese in
this compound may be in an intermediate spin state.\cite{prl} This
would significantly modify our understanding of the origin of CMR
in this compound in terms of on-site Hund's rule coupling. We hope
that this work will motivate further studies on the electron doped
phase of rare-earth manganites.

The authors would like to acknowledge Peter Oppeneer, F. Steglich,
Andy Mackenzie, S.~K. Dhar and R. Pinto for helpful discussions
and encouragement. We would also like to thank Chris Muirhead,
Radoslav Chakalov and Peter Mann for experimental help. P.R.
wishes to thank the Leverhulme Trust for financial support. The
research of Z.H. and L.H.T. is supported by the Deutsche
Forschungsgemeinschaft through SFB 608.

\end{document}